\documentclass[12pt]{article}

\usepackage[utf8]{inputenc}

\usepackage[T1]{fontenc}

\usepackage[top=2.1cm, bottom=1.8cm, left=2.3cm, right=2cm]{geometry}
\usepackage{amsmath,amsfonts,amssymb,amsthm}
\usepackage{mathrsfs}  
\usepackage{soul}
\usepackage{yhmath}
\usepackage{pgf, tikz}
\usepackage{lastpage}
\usepackage{pstricks}
\usepackage{wasysym}
\usepackage{pxfonts}

\usepackage[hidelinks]{hyperref}

\usepackage[toc]{appendix}

\usepackage{biblatex} 
\newcommand{\deriv}{\mathrm{d}}

\newcommand{\dvol}{\deriv\mathrm{Vol}^4}
\newcommand{\N}{\mathbb{N}}

\newcommand{\Err}{\mathrm{Err}}

\newtheorem{theorem}{Theorem}[section]
\newtheorem{proposition}{Proposition}[section]

\newtheorem{lemma}{Lemma}[section]

\begin{document}
\begin{center}
{\Large {\bf Peeling-off behaviour of the wave equation on the Vaidya spacetime}}
\vspace{0.3in}

{\large Armand Coudray\footnote{Laboratoire de Math\'ematiques de Bretagne Atlantique, UMR CNRS 6205, 6 avenue Le Gorgeu, 29238 Brest cedex 3, France}}
\end{center}
\tableofcontents

\section{Introduction}
\label{introduction}

In 1961, Sachs studied outgoing radiation fields from a source in curved space-times along null geodesics. In \cite{Sachs:1961zz}, he expanded the Riemann tensor in negative powers of $r$, with $r$ an affine parameter along outgoing null geodesics. 
The Sachs \textit{peeling property} of a field can be described as the alignment of its principal null directions along the generator of the geodesics when moving away from the source of radiation. 

In 1965, Penrose proved in \cite{Penrose:1965am}, that the peeling property can be reinterpreted in much simpler terms using a conformal compactification. Penrose's conformal methods consist in embedding a physical spacetime $(\mathcal{M},g)$ into a larger "compactified" spacetime $(\hat{\mathcal{M}}, \hat{g})$, where $\mathcal{M}$ is the interior of $\hat{\mathcal{M}}$ and 

 \begin{equation*}
     \hat{g} = \Omega^2 g; 
 \end{equation*}
 
the conformal factor $\Omega$  being a defining function of the boundary of $\hat M$ : $\partial \hat{\mathcal{M}} = \hat{\mathcal{M}} - \mathcal{M}$. This boundary is composed of two null hypersurfaces $\mathscr{I}^{\pm}$ referred to as future and past null infinities. Penrose showed that the Sachs peeling property is equivalent to the continuity of the conformal field at the boundary.

Then comes the question of how large a class of initial data ensures the peeling behaviour. On Minkowski spacetime, there is a simple answer based on a complete compactification in which the rescaled spacetime really is compact. As soon as the spacetime contains energy, there is singular at spacelike infinity and the question becomes difficult : if the field does not decay sufficiently fast at spacelike infinity, a singularity could creep up null infinity and make the whole asymptotic behaviour singular.

In 2007, Lionel Mason and Jean-Philippe Nicolas gave a complete description of peeling for scalar field on Schwarzschild's spacetime in \cite{mason_nicolas_2009}.

They established that the same assumptions on regularity and decay of initial data as in Minkowski's spacetime entailed the same regularity at null infinity of the conformal field. In \cite{2012JGP....62..867M} they extended their results to Dirac and Maxwell fields in Schwarzschild space-time. Then Pham and Nicolas obtained similar theorems for linear and semilinear scalar fields on Kerr metrics in \cite{Nicolas2019PeelingOK}. 

The peeling property defined by Mason and Nicolas is little bit different from the Penrose version because regularity at infinity is characterised by Sobolev norms instead of $C^k$ spaces. More precisely the functions spaces on a Cauchy hypersurface and null infinity are obtained from energy norms associated to well chosen vector fields. The idea is to obtain an equivalence at all orders between the energy of initial data and the energy at the boundary. Then they compared regularity and decay assumptions with the corresponding ones on Minkowski's spacetime and show that they are equivalent.

 Has alluded to above, the singularity at $i_0$ is directly linked to the black hole mass. The peeling of massless field as so far only been studied in stationary situations. Does the property still hold when the mass of the black hole varies with time? We study this question for Vaidya metrics. Vaidya black holes are spherically symmetric and have a varying mass due to the absorption or the emission of null dust that is transported along null geodesics. In this article we focus on the case of a white hole that evaporates via the emission of null dust.  

In \cite{Coudray2021GeometryOV} Jean-Philippe Nicolas and the author studied the geometry of such spacetimes with a particular emphasis on spherically symmetric optical functions and the corresponding null geodesic congruences. The second optical function (analogous to the advanced time of Eddington-Finkelstein coordinates on Schwarzschild's spacetime) will be useful  to us in controlling the geometry of the spacetime near $i_0$ and $\mathscr{I}$. 

For a completely general mass function, we characterise entirely the peeling property of massless scalar field and we prove that conditions for peeling at any given order are equivalent to those on flat spacetime.

The paper is organised as follow :  
 \begin{enumerate}
     \item We recall the definition and properties of Vaidya's spacetime, its conformal compactification and the wave equation on this background. In the remainder of the paper we work only on the compactified spacetime. 
     
     \item We present then the vector field method for the wave equation. We extended to Vaidya's spacetime the definition of the so-call Morawetz vector field and after a choice of stress-energy tensor, we give the expression of energy current and its divergence. This divergence is non-zero due to the fact that the Morawetz vector field is only an asymptotic Killing vector and the stress-energy tensor is not exactly conserved. 
     We established that the divergence terms are controlled by the energy of the field on the leaves of a foliation that we construct. 
     
     We obtain a first theorem that give a "peeling at order 0".   
     \item We obtain similar results for higher order energies. The main difficulty here is linked to the non-stationarity of Vaidya's spacetime. We establish the peeling at all orders and observe some subtle differences compared to the Schwarzschild case.  
 \end{enumerate}

\section{Geometrical setting}
\label{geometrical_setting}

\subsection{Construction of Vaidya's spacetime}

The Schwarzschild metric describes a static space-time with a spherical, isolated black hole of constant mass $M$; in spherical coordinates $(t,r,\theta,\phi)$ it is given by :

\begin{equation}
    g_{\mathrm{Sch}} = F(r) \deriv t^2 - F(r)^{-1} \deriv r^2 - r^2 \deriv \omega^2 \label{g_sch_t-r}
\end{equation}

with :

\begin{equation*}
    F(r) =1- \dfrac{2M}{r}, \deriv \omega^2 = \deriv \theta^2 + \sin^2\theta \deriv \phi^2 
\end{equation*}

The metric has a curvature singularity at $r=0$. The locus $r=2M$ is a fictitious singularity that can be understood as the union of two null hypersurfaces (the future and the past event horizon) by means of Eddington-Finkelstein coordinates. Outgoing Eddington-Finkelstein coordinates $(u,r,\theta, \phi)$ are defined by $u=t-r_{\star}$, with

\begin{equation}
   r_{\star} = r + 2M\log\left(r - 2M\right),\hspace{1cm}  \deriv r_{\star} = \dfrac{\deriv r}{F(r)}  
\end{equation}

In these coordinates the Schwarzschild metric reads 

\begin{equation}
    g_{\mathrm{Sch}} = F\deriv u^2 + 2 \deriv u \deriv r - r^2 \deriv \omega \label{Scwharzschild-outgoing}
\end{equation}

The Vaidya metric is defined from \eqref{Scwharzschild-outgoing} by allowing the mass $m$ to depend of the retarded time $u$.

\begin{equation}
    g = F(u,r) \deriv u^2 + 2 \deriv u \deriv r - r^2 \deriv \omega^2, \hspace{0.8cm} F(u,r) = 1 - \dfrac{2m(u)}{r} \label{vaidya-outgoing}
\end{equation}

Alternatively we can construct the Vaidya metric using ingoing Eddington-Finkelstein coordinates $(v,r,\theta, \phi)$ with $v = t + r_{\star}$. 

\begin{equation}
    g = F(v,r) \deriv v^2 - 2 \deriv v \deriv r - r^2 \deriv \omega^2, \hspace{0.8cm} F(v,r) = 1 - \dfrac{2m(v)}{r} \label{vaidya-ingoing}
\end{equation}

Metrics \eqref{vaidya-outgoing} and \eqref{vaidya-ingoing} are solutions of Einstein equations with a source. \eqref{vaidya-outgoing} describes a white hole that evaporates classically via the emission of null dust whereas \eqref{vaidya-ingoing} corresponds to a black hole that mass increases as a result of accretion of null dust.

In this paper, we will deal with an evaporating white hole, hence the mass $m$ is a decreasing function of $u$. We assume that the mass is a smooth decreasing function of $u$.

Conformal compactification of Vaidya spacetime, with conformal factor $\Omega = R = 1/r$ :

\begin{equation}
    \hat{g} = R^2 F(u,R) \deriv u^2 - 2 \deriv u \deriv R - \deriv\omega, \hspace{0.8cm} F(u,R) = 1 - 2m(u)R. 
\end{equation}

The inverse rescaled metric is  :

\begin{equation}
    \hat{g}^{-1} = -R^2 F(u,R) \partial_R^2 - 2 \partial_R \partial_u - \partial^2_{\omega^2}
\end{equation}

The d'Alembertien associated to this compactified metric
\[ \square_{\hat{g}} := \nabla_a \nabla^a \, ,\]
where $\nabla$ is the Levi-Civita connection associated to $\hat{g}$, is given in terms of coordinates ($u,R,\omega$) by :
\begin{equation}
    \square_{\hat{g}} = -2\partial_u \partial_R - \partial_R R^2(1-2m(u)R)\partial_R - \Delta_{S^2} \label{d_alembertien_wave-operator}
\end{equation}

The Ricci scalar, is defined by $R^{ab}R_{ab}$ i.e. the contraction of the Ricci tensor with itself :  

\begin{equation*}
    \mathrm{Scal}_{\hat{g}} = R^{ab}R_{ab} = 12m(u)R \;, ~  \mathrm{Scal}_{g} = 0.
\end{equation*}

The metric \eqref{vaidya-outgoing} has the following non-zero Christoffel symbols, with $m'(u) = \deriv m/\deriv u$ :

\begin{gather*}
\Gamma_{ 0 \, 0 }^{ 0  }  =  -3R^2 m(u) + R \, ,~ \Gamma_{ 0 \, 1 }^{ 1  }  =  3R^2 m(u) - R \, , \\
\Gamma_{ 0 \, {0} }^{ \, 1}  =  6R^5 m^2(u) - 5 R^4 m(u) + R^3 m'(u) + R^3  \,   , \\
\Gamma_{ 3 \, 3 }^{ 2 }  =  -\cos\theta\sin\theta \, ,~ \Gamma_{ 3 \, 2 }^{ 3 } =  \frac{\cos \theta}{\sin \theta} \, .
\end{gather*}

\subsection{The second optical function on Vaidya space-time}

We can construct a second optical function analogous to $v= t+ r_{\star}$ in the Schwarzschild spacetime. This has been done in details in \cite{Coudray2021GeometryOV}. 

On the Schwarzschild metric we have $g = F\deriv u \deriv v - r^2 \deriv \omega^2$; for Vaidya's spacetime we have 

\begin{equation*}
    g = F \deriv u (\deriv u + \dfrac{2}{F}\deriv r) - r^2 \deriv \omega^2
\end{equation*}

and the 1-form $ \deriv u + \dfrac{2}{F}\deriv r$ is not exact. However, introducing an auxiliary positive function $\varphi$ we can write :

\begin{equation}
g = \dfrac{F}{\psi}\deriv u \left(\psi \deriv u +  2 \psi F^{-1} \deriv r\right) \label{metric_varphi} - r^2 \deriv \omega^2
\end{equation}

and arrange for the 1-form, $\psi \deriv u +  2 \psi F^{-1} \deriv r$ to be exact and null, if we assume :

\begin{equation} \label{EDOphi}
    \dfrac{\partial \psi}{\partial u} - \dfrac{F(u,r)}{2}\dfrac{\partial\psi}{\partial r} + \dfrac{2m'(u)}{Fr} \psi = 0 \, .
\end{equation}

This equation can be solved by setting, say, $\psi = 1$ on $\mathscr{I}^{-}$ and integrating along incoming principal null geodesics (see \cite{Coudray2021GeometryOV}). Then we define $v$ by 

\begin{equation*}
    \deriv v= \psi \, \deriv u + 2 \frac{\psi}{F} \, \deriv r \label{deriv_v}
\end{equation*}

and $v = -\infty$ on $\mathscr{I}^{-}$. 

It is useful to define new radial and time variables such that :
\[
\left\{ \begin{array}{l} {t = u + \tilde{r}\, ,}\\
{t = v - \tilde{r}\, .} \end{array} \right.
\]

The relations between their differentials are 

\begin{align}
      \deriv \Tilde{r} =& \dfrac{1}{2}\left(\psi - 1 \right) \deriv u + \frac{\psi}{F} \deriv r \label{deriv_r_tilde}\\
      \deriv t = & \dfrac{1}{2}\left( \psi + 1 \right) \deriv u + \frac{\psi}{F} \deriv r \label{deriv_t}
\end{align}

\section{Vector field method}
\label{vector-field_method}

Let $\Psi$ be a solution to the equation 

\begin{equation}
    (\square_{g} + \frac{1}{6}\mathrm{Scal}_g) \Psi = 0, \label{equation_onde_physique}
\end{equation}

This is equivalent to $\phi = \Omega^{-1} \Psi$ solution to :

\begin{equation}
    (\square_{\hat{g}} + \frac{1}{6}\mathrm{Scal}_{\hat{g}}) \phi = 0, \label{equation_ondes}
\end{equation}

We will study the peeling for solutions to \eqref{equation_onde_physique} entirely at the level of the rescaled field $\phi$ by performing geometric energy estimates for solutions to \eqref{equation_ondes}. This can be obtained in the following manner :

\begin{enumerate}
    \item choose a stress-energy tensor $T_{ab}$ and a causal vector field (the observer) $T^a$;
    \item contract these two quantities to compute the associated energy current; 
    \item use the divergence Theorem on a closed hypersurface. 
\end{enumerate}

The Energy flux of a field $\phi$ through a hypersurface $S$ is defined by :

\begin{equation*}
    \mathcal{E}_{S}(\phi) = \int_{S} \star T^a T_{ab}
\end{equation*}

where $\star$ denotes the Hodge dual. We choose the energy momentum tensor as : 

\begin{equation}
    T_{ab} = \nabla_a \phi \nabla_b \phi - \dfrac{1}{2}\hat{g}_{ab}\nabla_c \phi \nabla^c \phi \label{def_tens_imp_ener}\, .
\end{equation}
We need an observer that is transverse to null infinity so as to have an energy on $\mathscr{I}^{+}$ that controls all tangential derivatives instead of merely $\partial \phi /\partial u$. We choose the so-called Morawetz vector field : 
\begin{equation}
    T = u^2 \partial_u - 2(1+uR)\partial_R
\end{equation}
We establish in Appendix \ref{AppMorawetz} that it is timelike and future-oriented on $\Omega_{u_0}$ (we choose $\partial_u$ as the global time orientation for our spacetime).

The corresponding energy current is given by the 3-form :
\begin{eqnarray*}
    \star T^a T_{ab} &=& T^a \nabla_a \star \nabla_b \phi -\dfrac{1}{2}\nabla_ c \phi \nabla^c \phi \star T_ b \\
 &=&\left[u^2 \phi_u^2 + R^2 F(u,R)(u^2 \phi_u \phi_R - (1+uR)\phi_R^2 ) + (1+uR)\vert \nabla_{S^2}\phi\vert^2 \right] \deriv u \wedge \deriv \omega\\
    && +\dfrac{1}{2}\left[\left((2+uR)^2 -2m(u)u^2 R^3 \right) \phi_R^2 + u^2 \vert \nabla_{S^2} \phi\vert^2 \right] \deriv R \wedge \deriv \omega
\end{eqnarray*}
Now to obtain the energy flux through an oriented hypersurface, we integrate on it the previous $3$-form.

\subsection{Hypersurfaces and identifying vector}

For $u_0 <<-1$ given, let us consider the neighbourhood of $i^0$
\[ \Omega_{u_0} := \{ t \geq 0 \} \cap \{ u < u_0 \} \, .\]
All our estimates will be established in $\Omega_{u_0}$ and $u_0$ will be chosen large enough in absolute value to ensure a finite number of basic inequalities, such as those given in Proposition \ref{prop_first_estimation}.
Note that since $t \geq 0$ on $\Omega_{u_0}$ and $u_0<<-1$, we necessarily have that $R <<1$ in this domain.

We define a foliation of $\Omega_\varepsilon$ by spacelike hypersurfaces $\mathcal{H}_s$:
\begin{equation}
    \mathcal{H}_s =\lbrace u = -s \tilde{r} \rbrace \cap \lbrace u < u_0\rbrace, ~0<s \leq 1 \, ,
\end{equation}
where $u_0 \ll -1$ is given. The co-normal 1-form on these hypersurfaces is given by:
\begin{equation*}
    \omega = \left[1 + \dfrac{s}{2}(\psi - 1)\right]\deriv u + s\frac{\psi}{F} \deriv r, \, .
\end{equation*}
One can easily verify that $g^{-1}(\omega , \omega ) >0$, whence the normal vector ($\hat{g}^{-1} \omega$) is a timelike and the hypersurfaces $\mathcal{H}_s$ are spacelike. The hypersurface $\mathcal{H}_1$ corresponds to $t=0$ and this is where we shall set our initial data. We can in addition define the hypersurface $\mathcal{H}_0$ as the limit of $\mathcal{H}_s$ as $s\rightarrow 0$; for fixed $u$, as $s \rightarrow 0$, we have $\tilde{r} \rightarrow +\infty$, so $\mathcal{H}_0$ is the set of points at infinity for which $u < u_0$,
\[ \mathcal{H}_0 = \mathscr{I}^+ \cap \{ u<u_0 \} =: \mathscr{I}^+_{u_0} \, .\]
We choose an identifying vector $\nu$ i.e. a vector that is transverse to all the surfaces of the foliation and such that it crosses each surface only once.
 \begin{equation} \label{nu}
    \nu  = \dfrac{\Tilde{r}^2 R^2 F}{\psi \vert u \vert}\partial_R
\end{equation}
And we have also on $\mathcal{H}_s$ :
\begin{equation}
\deriv u \vert_{\mathcal{H}_s}= \left.\dfrac{2s\psi}{F\left[2 + s(\psi  -1)\right]}\dfrac{\deriv R}{R^2}\right\vert_{\mathcal{H}_s}, \;\;\deriv R\vert_{\mathcal{H}_s} = \dfrac{FR^2}{s \psi}\left[1 + \frac{s}{2}(\psi - 1)\right] \deriv u \vert_{\mathcal{H}_s} \label{dR/du_H_s}
\end{equation}

The boundary of $\Omega_{u_0}$ is made of the hypersurfaces $\mathcal{H}_0=\mathscr{I}^+_{u_0}$, $\mathcal{H}_1$ and a third one defined by
\begin{equation}
\mathcal{S}_{u_0} := \{ u=u_0 \} \cap \{ t \geq 0 \} \, .
\end{equation}.

\subsection{Energy fluxes and estimates}

The expression of energy fluxes across the hypersurfaces $\mathcal{H}_s, \mathcal{H}_0$ and $\mathcal{S}_u$ is as follows
\begin{align*}
    \mathcal{E}_{\mathcal{H}_s}(\phi) =& \int_{\mathcal{H}_s} \left\lbrace u^2 \phi_u^2 + u^2 R^2 F(u,R) \phi_u \phi_R + \vert \nabla_{S^2}\phi\vert^2\left[u^2\dfrac{FR^2}{4s \psi}\left(2 + s(\psi - 1)\right) + (1+uR) \right] \right.\\
    & \left. + R^2 F  \left[\dfrac{2 + s(\psi - 1)}{4s \psi}\left( (2+uR)^2 - 2 m(u)u^2R^3\right) - (1+uR)\right]\phi_R^2 \right\rbrace \deriv u \wedge \deriv \omega \vert_{\mathcal{H}_s}\\
    \mathcal{E}_{\mathscr{I}^+_{u_0}}(\phi) =& \int_{\mathcal{H}_{0}} \left[ u^2 \phi_u^2 + \vert \nabla_{S^2}\phi \vert^2 \right]\deriv u \wedge \deriv \omega\\
    \mathcal{E}_{\mathcal{S}_u}(\phi) = & \int_{\mathcal{S}_u} \dfrac{1}{2}\left[\left( (2+uR)^2 -2m(u) u^2 R^3\right) \phi_R^2 + u^2 \vert \nabla_{S^2} \phi \vert^2 \right] \deriv R \wedge \deriv \omega \label{E_S_u}
\end{align*}
These expressions are rather complicated but in $\Omega_{u_0}$ they can be simplified to obtain more usable approximate forms
\begin{proposition}
\label{prop_first_estimation}
Let $\varepsilon>0$, then for $u_0<<-1$ large enough in absolute value, we have in $\Omega_{u_0}$: 
\begin{gather*}
    1\leq \psi <1+\varepsilon , \; 1-\varepsilon <\tilde{r}R<1+\varepsilon , \;\\
    0\leq R\vert u \vert<1+\varepsilon, \; 1-\varepsilon< 1-2m(u)R\leq 1 .
\end{gather*}
\end{proposition}
The proof is given in Appendix \ref{AppProofs} Section \ref{ProofProp31}.

Using this we can compute an equivalent form of the energy flux across $\mathcal{H}_s$. 
\begin{proposition}
\label{prop_equivalence}
In $\Omega_{u_0}$, we have the equivalence:
\begin{equation}
    \mathcal{E}_{\mathcal{H}_s}(\phi) \simeq \int_{\mathcal{H}_s} \left[u^2\vert \partial_u\phi\vert^2  + \dfrac{R}{\vert u \vert}\vert\partial_R\phi\vert^2 + \vert \nabla_{S^2} \phi\vert^2 \right]\deriv u \wedge \deriv \omega \vert_{\mathcal{H}_s} \label{energy_H_s}
\end{equation}

The proof is given in Appendix \ref{AppProofs} Section \ref{ProofProp32}. 
\end{proposition}

\subsection{Error terms and Stokes' Theorem} 
Our approach to the Peeling is based on energy estimates. Stokes' Theorem will allow us to obtain inequalities between the energy fluxes through our different hypersurfaces, from a conservation law for the energy current. However, we have to be careful because the conservation law is only approximate. There will be two types of error terms coming from the non-zero divergence of the stress-energy tensor and from the fact that our observer is not a Killing vector field:
\begin{equation*}
    \nabla^{(a} \left(T^{b)}T_{ab}\right) = \nabla^{(a} T^{b)} T_{ab} + T^b \nabla^a T_{ab} \, .
\end{equation*}
The divergence of \eqref{def_tens_imp_ener}, provided $\phi$ satisfies equation \eqref{equation_ondes}, is given by, 

\begin{equation} \label{divTab}
    \nabla^{a}T_{ab} = \square_{\hat{g}} \phi \nabla_b \phi = - 2m(u) R \phi \partial_b \phi \, .
\end{equation}
The Morawetz vector field is not Killing but its  Killing form (or deformation tensor) tends to zero at infinity: 
\begin{equation}
   \nabla^{(a}T^{b)} = \left( -u^2 R^3 m'(u) + 2mR^2 (3+uR) \right)\partial_R^a \partial_R^b \label{nabla^a_T^b}
\end{equation}
Given $0 \leq s_1 \leq s_2 \leq 1$, we define the domain
\[ \Omega_{u_0,s_1,s_2} = \Omega_{u_0} \cap \{ s_1 \leq s \leq s_2 \} \]
and the part of $S_{u_0}$ that is in the boundary of $\Omega_{u_0,s_1,s_2}$:
\[ S_{u_0,s_1,s_2} = S_{u_0} \cap \{ s_1 \leq s \leq s_2 \} \, .\]
The equations \eqref{divTab} and \eqref{nabla^a_T^b} as well as Stokes' Theorem give us the following identity for any scalar field $\phi$ on $\Omega_{u_0}$
\begin{gather} 
\mathcal{E}_{\mathcal{H}_{s_2}} (\phi) + \mathcal{E}_{S_{u_0,s_1,s_2}} (\phi) - \mathcal{E}_{\mathcal{H}_{s_1}} (\phi) \hspace{3in} \nonumber \\
\hspace{1.5in} = \int_{\Omega_{u_0,s_1,s_2}} \left( \square_{\hat{g}} \phi \partial_T \phi + \left( -u^2 R^3 m'(u) + 2mR^2 (3+uR) \right) T_{11} (\phi ) \right) \mathrm{dVol}^4 \, .\label{EnId}
\end{gather}
The right-hand side of \eqref{EnId} can be decomposed into an integral in $s$ over $[s_1,s_2]$ of integrals over $\mathcal{H}_s$; this is done by splitting the $4$-volume measure using the identifying vector field $\nu$ (see \eqref{nu}) as follows
\[ \mathrm{dVol}^4 = \deriv s \wedge \nu\lrcorner \mathrm{dVol}^4 \]
and
\[ \deriv s \wedge \nu\lrcorner \mathrm{dVol}^4 \vert_{\mathcal{H}_s} = \dfrac{(\tilde{r} R)^2}{\varphi \vert u \vert}\deriv u\wedge \deriv \omega \, .\]
Equation \eqref{EnId} then becomes
\begin{gather} 
\mathcal{E}_{\mathcal{H}_{s_2}} (\phi) + \mathcal{E}_{S_{u_0,s_1,s_2}} (\phi) - \mathcal{E}_{\mathcal{H}_{s_1}} (\phi) \hspace{3in} \nonumber \\
\hspace{1.5in} = \int_{s_1}^{s_2} \left( \int_{\mathcal{H}_{s}} \Err (\phi ) \deriv u \wedge \deriv \omega \right) \deriv s\, , \label{conservation}
\end{gather}
where
\begin{equation} \label{GenErrTerm}
\Err (\phi ) = \bigg( \square_{\hat{g}} \phi \partial_T \phi + \left( -u^2 R^3 m'(u) + 2mR^2 (3+uR) \right) T_{11} (\phi ) \bigg) \dfrac{(\tilde{r} R)^2}{\varphi \vert u \vert} \, .
\end{equation}
In the case where $\phi$ satisfies equation \eqref{equation_ondes}, the error term \eqref{GenErrTerm} becomes
\begin{equation} \label{ErrTermPhiFinalForm}
    \Err(\phi)= \left[\left( -u^2 R^3 m'(u) + 2mR^2 (3+uR) \right) \phi_R^2 -2mR\phi \left( u^2 \phi_u - 2(1+uR)\phi_R\right)\right] \dfrac{(\tilde{r} R)^2}{\varphi \vert u \vert}
\end{equation}

where $\phi_u$ and $\phi_R$ are respectively $\partial_u \phi$ et $\partial_R \phi$.

\subsection{Control of the error terms}

In this subsection we focus on the control of the error terms. We show that the error term is almost entirely controlled by the energy density on $\mathcal{H}_s$, except for, a priori, a additional $L^2$ term.
\begin{proposition}
\label{prop_error_term}
For $u<u_0$, $u_0\ll -1$ and $R\to 0$ :
\begin{equation*}
  \vert \Err(\phi) \vert \lesssim u^2 \phi_u^2 + \dfrac{R}{\vert u \vert} \phi_R^2 + \phi^2
\end{equation*}
\end{proposition}
The proof is given in Appendix \ref{AppProofs} Section \ref{ProofProp33}. Then we use a Poincaré-type estimate proved in \cite{mason_nicolas_2009}, in order to control the additional term.
\begin{lemma}
\label{lemma_L2_H}
For $u_0<0$, there exists $C>0$ , in $\mathbb{R}$ such that for all bounded support function $f\in \mathcal{C}^{\infty}_0(\mathbb{R})$ we have :
\begin{equation*}
    \int_{-\infty}^{u_0} (f(u))^2 \deriv u \leq C \int^{u_0}_{-\infty}u^2 (f'(u))^2 \deriv u
\end{equation*}
with $f'(u) = \deriv f /\deriv u$. And an immediate consequence is that for $0\leq s \leq 1$ :
\begin{equation*}
    \int_{\mathcal{H}_{s,u_0}} \phi^2 \deriv u \deriv \omega \lesssim \mathcal{E}_{\mathcal{H}_s}(\phi)
\end{equation*}
\end{lemma}
This lemma together with Propositions \ref{prop_equivalence} and \ref{prop_error_term} entail:
\begin{theorem}
\label{thm_error-terms}
In the domain $\Omega_{u_0}$, $u_0<-1$  large enough in absolute value, we have :
\begin{equation*}
  \int_{\mathcal{H}_s} \Err(\phi)\; \deriv u \deriv \omega \vert_{\mathcal{H}_s}\lesssim  \; \mathcal{E}_{\mathcal{H}_s}(\phi) 
\end{equation*}
\end{theorem}

\section{Peeling}
\label{peeling_section}

To obtain a peeling theory in Vaidya spacetime, we use the approximate conservation law \eqref{conservation} for $\phi$ solution to \eqref{equation_ondes} and its successive derivatives, in order to obtain estimates both ways between the energy on $\mathscr{I}^+$ and $\mathcal{H}_1$ at all orders of regularity. This is done as follows:

\begin{enumerate}
    \item we obtain the fundamental estimate for $\phi$ using the control of the error terms and a Grönwall inequality;
    \item then we work out the equations satisfied by $\phi_R$, $\phi_u$, $\partial_R^k \partial_u^l \phi$ and follow the same procedure to infer higher order estimates.
\end{enumerate}

\subsection{Fundamental estimates}
\label{fundamental estimations}

\begin{theorem}
\label{theorem_fundamental_estim}
For $u_0 \ll -1$, $0 \leq s_0 \leq 1$, for any $\phi$ solution to \eqref{equation_ondes} associated with initial data in $\mathcal{C}^{\infty}_0$, we have :
\begin{align}
    \mathcal{E}_{\mathcal{H}_{s}}(\phi) \lesssim &\;\;\mathcal{E}_{\mathcal{H}_{1}}(\phi) \, , \label{BEE1}\\
    \mathcal{E}_{\mathcal{H}_{s}}(\phi) \lesssim &\;\;\mathcal{E}_{\mathscr{I}_{u_0}^{+}}(\phi) + \mathcal{E}_{S_{u_0,0,s}}(\phi) \, .\label{BEE2}
\end{align}
For $s=0$, \eqref{BEE1} becomes: 
\begin{equation*}
    \mathcal{E}_{\mathcal{H}_{\mathscr{I}^+_{u_0}}}(\phi) \lesssim \mathcal{E}_{\mathcal{H}_1}(\phi)
\end{equation*}
and for $s = 1$, \eqref{BEE2} gives
\begin{equation*}
      \mathcal{E}_{\mathcal{H}_{1}}(\phi) \lesssim \mathcal{E}_{\mathscr{I}_{u_0}^{+}}(\phi) + \mathcal{E}_{S_{u_0}}(\phi) \, .
\end{equation*}
\end{theorem}

\textbf{Proof.} Conservation law \eqref{conservation} entails
\begin{equation}
    \mathcal{E}_{\mathcal{H}_s}(\phi) + \mathcal{E}_{\mathcal{S}_{u_0,s,1}}(\phi) - \mathcal{E}_{\mathcal{H}_1}(\phi) \leq \int^{s}_1 \int_{\mathcal{H}_{\tilde{s}}} \vert\Err(\phi) \vert \deriv u \deriv \omega \deriv \tilde{s} \, .
\end{equation}
Before continuing the demonstration we have to prove that on surfaces $\mathcal{S}_u$  we have a non-negative energy. As a consequence of the estimates of Proposition \ref{prop_first_estimation}, we have in $\Omega_{u_0}$: 
\begin{equation*}
    (2 +uR)^2 -2m(u)u^2 R^3 \simeq (2 + uR)^2 > 0
\end{equation*}
and the density of energy on $S_{u_0}$ is therefore non-negative. Note that this can also be inferred geometrically, since the Morawetz vector field $T^a$ is timelike and future-oriented on the whole of $\Omega_{u_0}$, $S_{u_0}$ is null and we choose the future-oriented normal $-\partial_R$ on it and finally the stress-energy tensor \eqref{def_tens_imp_ener} satisfies the dominant energy condition.

So \eqref{conservation} yields:
\begin{equation}
    \mathcal{E}_{\mathcal{H}_s}(\phi) - \mathcal{E}_{\mathcal{H}_1}(\phi) \leq \int^{s}_1 \int_{\mathcal{H}_{\tilde{s}}} \vert\Err(\phi) \vert \deriv u \deriv \omega \deriv \tilde{s} \, .
\end{equation}
Theorem \ref{thm_error-terms} allows to control the error term by the energy density on $\mathcal{H}_{\tilde{s}}$ and so:

\begin{equation}
    \mathcal{E}_{\mathcal{H}_s}(\phi) - \mathcal{E}_{\mathcal{H}_1}(\phi) \leq \int^{s}_1  \mathcal{E}_{\mathcal{H}_{\tilde{s}}}(\phi)  \deriv \tilde{s} \, .
\end{equation}
Grönwall's lemma then gives equation \eqref{BEE1}.

The proof of \eqref{BEE2} is similar. We have
\begin{align*}
     \mathcal{E}_{\mathcal{H}_{s}}(\phi) \leq & \; \mathcal{E}_{\mathscr{I}^{+}_{u_0}}(\phi) + \mathcal{E}_{\mathcal{S}_{u_0,0,s}}(\phi) +\int^{s}_{0} \int_{\mathcal{H}_{\tilde{s}}} \vert\Err(\phi)\vert \deriv u \deriv \omega \deriv \tilde{s} \, ,\\
      \mathcal{E}_{\mathcal{H}_{s}}(\phi) \lesssim & \; \mathcal{E}_{\mathscr{I}^{+}_{u_0}}(\phi) + \mathcal{E}_{\mathcal{S}_{u_0,0,s}}(\phi) + \int^{s}_{0} \mathcal{E}_{\mathcal{H}_{\tilde{s}}}(\phi) \deriv \tilde{s} \, .
\end{align*}

and we conclude using Grönwall's lemma. \qed

\subsection{Higher order estimates}
\label{high order}

Theorem \ref{theorem_fundamental_estim} gives us the basic estimates both ways between $\mathscr{I}^+$ and the initial data hypersurface. In order to prove estimates for derivatives of the solution, we commute partial derivatives into Equation \ref{equation_ondes}. The Vaidya metric is not stationary and this means that $\partial_u$ does not commute with the wave equation. As a consequence, a control of $\partial_u \phi$ will require a joint control of $\partial_R \phi$ because of the relations:
\begin{eqnarray}
\left[ \partial_u , \square_{\hat{g}} \right] &=&  2m'(u) R^3 \partial_R^2 + 6m'(u) R^2 \partial_R \, , \label{derivee_laplacien_u} \\
\left[ \partial_u , \mathrm{Scal}_{\hat{g}} \right] &=& 12 m'(u) R \, .
\label{derivee_scal_u}
\end{eqnarray}
By contrast, when commuting the $R$ derivative with the d'Alembertian, we only obtain error terms involving $\partial_R$:
\begin{eqnarray}
    \left[ \partial_R , \square_{\hat{g}} \right] &= & -2R(1-3mR) \partial_R^2 - 2(1-6mR)\partial_R \, , \label{derivee_laplacien_R} \\
    \left[ \partial_R , \mathrm{Scal}_{\hat{g}} \right] &=& 12 m(u) \, .
    \label{derivee_scal_R}
\end{eqnarray}
Hence, just as in the Schwarzschild case, we can control the successive derivatives with respect to $R$ independently of the other variables. However, the energy of $\partial_R^k \partial_u^l \phi$ will need to be controlled by those of $\partial_R^p \partial_u^q \phi$ with $p+q\leq l+k$ and $k\leq p \leq k+1$. We obtain the following result. 
\begin{theorem}
\label{thm_peeling_ho}
We have the following inequalities:
\begin{enumerate}
\item The $R$ derivatives, are controlled independently of the others, just like angular derivatives: for all $k \in \N$,
\begin{align*}
\mathcal{E}_{\mathscr{I}^{+}_{u_0}}\left(\partial_{R}^{k} \phi\right) \lesssim & \mathcal{E}_{\mathcal{H}_1}(\partial_R^{k} \phi ) \, , \\
\mathcal{E}_{\mathcal{H}_{1,u_0}}\left(\partial_R^{k} \phi \right) \lesssim & \mathcal{E}_{\mathscr{I}^{+}_{u_0}}(\partial_R^{k} \phi) + \mathcal{E}_{\mathcal{S}_{u_0}}(\partial_R^{k} \phi)  \, , \\
\mathcal{E}_{\mathscr{I}^{+}_{u_0}}\left(\nabla_{S^2}^{k} \phi\right) \lesssim & \mathcal{E}_{\mathcal{H}_1}(\nabla_{S^2}^{k} \phi ) \, , \\
\mathcal{E}_{\mathcal{H}_{1,u_0}}\left(\nabla_{S^2}^{k} \phi \right) \lesssim & \mathcal{E}_{\mathscr{I}^{+}_{u_0}}(\nabla_{S^2}^{k} \phi) + \mathcal{E}_{\mathcal{S}_{u_0}}(\nabla_{S^2}^{k} \phi)  \, .
 \end{align*}
    \item The general control on partial derivatives of all orders has the following form: for all $k,l,n \in N$,
\begin{align*}
\mathcal{E}_{\mathscr{I}^{+}_{u_0}}\left(\partial_{R}^{k} \partial_u^l \nabla_{S^2}^{n} \phi\right) \lesssim & \sum_{p+q \leq k+l} \mathcal{E}_{\mathcal{H}_1}(\partial_R^{p}\partial_u^{q} \nabla_{S^2}^{n}\phi) \, , \\
\mathcal{E}_{\mathcal{H}_{1,u_0}}\left(\partial_{R}^{k} \partial_u^l \nabla_{S^2}^{n} \phi \right) \lesssim & \sum_{p+q \leq k+l} \left[\mathcal{E}_{\mathscr{I}^{+}_{u_0}}(\partial_R^{p}\partial_u^{q} \nabla_{S^2}^{n} \phi) + \mathcal{E}_{\mathcal{S}_{u_0}}(\partial_R^{p}\partial_u^{q} \nabla_{S^2}^{n} \phi) \right] \, .
 \end{align*}
 \end{enumerate}
\end{theorem}
The proof is given in Appendix \ref{AppendixProofThmPeeling}.

\newpage


\begin{appendix}
\section{On the Morawetz vector field} \label{AppMorawetz}

In this section, we prove that the Morawetz vector field is timelike and future-oriented in the neighbourhood of spacelike infinity and therefore transverse to null infinity. This allows us to define a positive or non-negative energy flux across spacelike or null hypersurfaces whose normals are future-oriented (for instance on $\mathcal{H}_s$, $\mathscr{I}^+$ or $\mathcal{S}_{u_0}$).

The ``squared norm'' of the vector field $T^a$ is given by:
\begin{equation*}
    \hat{g}(T,T) = u^2 \left[ F \vert uR\vert^2 - 4 \vert uR\vert + 4\right] \, .
\end{equation*}
The expression between square brackets admits two roots denoted by $\vert uR\vert_{\pm}$ : 
\begin{equation} \label{RouteS}
    \vert ur \vert_{\pm} = 2\dfrac{1 \pm \sqrt{2m(u)R}}{1-2m(u)R} \, .
\end{equation}
As one approaches $i^0$, both roots \eqref{RouteS} tend to 2 and we have seen that in $\Omega_{u_0}$, $0<\vert u \vert R<1+\varepsilon$. So this means that $\hat{g}(T,T)>0$, hence $T$ is timelike, in $\Omega_{u_0}$. Now choosing $\partial_u$ as the global time orientation of our spacetime, we have:
\begin{equation*}
    g(T,\partial_u) = \vert uR \vert^2 F - 2\vert uR \vert + 2 
\end{equation*}
and this is positive near $i^0$. The Morawetz vector field is therefore timelike and future oriented in $\Omega_{u_0}$. \qed

\section{Proof of main propositions and theorem} \label{AppProofs}

\subsection{Proof of proposition \ref{prop_first_estimation}} \label{ProofProp31}

Firstly we want to prove that in $\Omega_{u_0}$, provided we take $\vert u_0 \vert$ large enough, we have
\begin{equation*}
    1\leq \psi<1+\varepsilon \, .
\end{equation*}
We need a good understanding of the behaviour of the function $\psi$ near $i^0$. To do so, we integrate the differential equation \eqref{EDOphi} along $v=$cst. curves parametrised by $u$. We start from $\mathscr{I}^-$ where we set $\psi \equiv 1$. Equation \eqref{EDOphi} gives
\begin{equation}
    \left(\log (\psi)\right)' = \dfrac{-2Rm'(u)}{F} \label{def_varphi'}
\end{equation}
using the fact that the total derivative of $\varphi $ along the curve is
\[ \frac{\deriv \psi}{\deriv u} = \frac{\partial \psi}{\partial u } - \frac{F}{2} \frac{\partial \psi}{\partial r} \, .\]
From \eqref{def_varphi'}, we obtain an expression for $\varphi F$~:
\begin{equation}
\psi =\exp\left[ - \int_{-\infty}^{u} \dfrac{2R m'(\mu)}{1 - 2mR}\deriv \mu \right] =: \exp\left[-\int_{-\infty}^u f(\mu, R)\deriv \mu\right] \, . \label{varphi}
\end{equation}
And since $m'(\mu)\leq 0$ in the exterior region, we have $f(\mu,R) \leq 0 $.  Within $\Omega_{u_0}$ the function $F$ is bounded below by $F_{min}$ is arbitrarily close to 1, and bounded above by 1. Hence we have the following estimates for $f$:

\begin{equation*}
    0\leq \vert f(u,R) \vert\leq -2\dfrac{m'(u) R}{F_{\mathrm{min}}} \leq  -m' (u) \, .
\end{equation*}

Moreover, the function $f$ tends to 0 at $\mathscr{I}^{+}$, because $R \to 0$. Since the mass function has the finite limit $m_-$ as $u\rightarrow -\infty$, it follows that $m'$ is integrable in the neighbourhood of $-\infty$. We can therefore use Lebesgue's dominated convergence Theorem to obtain, uniformly for $u \leq u_0$
\begin{gather*}
    \lim \limits_{\mathscr{I^{+}}} \int_{-\infty}^u f(\mu, R) \deriv \mu = 0 \, , \\
    \lim \limits_{\mathscr{I^{+}}} (\psi) = \lim \limits_{\mathscr{I^{+}}} \exp\left[\int_{-\infty}^{u}f(\mu,R)\deriv \mu\right] = 1 \, .
\end{gather*}

Given $\varepsilon > 0$, Equation \eqref{varphi} entails that for $u_0 <0$, $\vert u_0 \vert$ large enough and $u\leq u_0$, we have

\begin{equation*}
    1 \leq \psi < 1+\varepsilon \, .
\end{equation*}
Now we want to prove that :
\begin{equation*}
    1-\varepsilon <\tilde{r}R<1+\varepsilon
\end{equation*}
Let $v_0 >>1$, we begin by integrating $r(u,v)$ between $v_0$ and $v$, on a $u=cst.$ line, using: 
\begin{equation*}
    \dfrac{\deriv r}{\deriv v} = \dfrac{F}{2\psi} \, ,
\end{equation*}
and $\tilde{r}= (v-u)/2$, so :
\begin{equation*}
    \dfrac{r(u,v)}{\tilde{r}(u,v)} = \dfrac{2}{v-u}\left(r(u,v_0) + \dfrac{1}{2}\int_{v_0}^v \dfrac{F}{\psi}\deriv v\right)
\end{equation*}
Taking $v_0<v_1 < v$ we have, 
\begin{equation}
    \dfrac{{r(u,v)}}{\tilde{r}(u,v)}-1 = \dfrac{2r(u,v_0) + u - v_1}{v-u} + \dfrac{1}{v-u}\int_{v_0}^{v_1}\dfrac{F}{\psi}\deriv v + \dfrac{1}{v-u}\int_{v_1}^v \left(\dfrac{F}{\psi}-1\right)\deriv v \label{rtildeR}
\end{equation}
At $v=v_0$ we have :
\begin{equation*}
   2r(u,v_0) = 2r(u_0,v_0) - \dfrac{1}{2}\int^{u}_{u_0} F\deriv u\,.
\end{equation*}
If we take $u<u_0$, with $u_0 \ll -1$, we know that $1-\varepsilon \leq F \leq 1$. Let also $v>v_0$ with $v_0$ large enough, 
\begin{align*}
    \dfrac{\varepsilon u + 2r(u_0,v_0) + u_0 - v_1}{v-u} &\leq\dfrac{2r(u,v_0) + u - v_1}{v-u} \leq \dfrac{2r(u_0,v_0) + u_0 - v_1}{v-u}\\
    -\dfrac{\varepsilon \vert u \vert}{v + \vert u \vert} & \leq \dfrac{2r(u,v_0) + u - v_1}{v-u} \leq \varepsilon\\
    -\varepsilon & \leq \dfrac{2r(u,v_0) + u - v_1}{v-u} \leq \varepsilon \, .
\end{align*}
The second term in \eqref{rtildeR} is also bounded by $\varepsilon>0$ near $i_0$ because $F/\psi \leq 1$ and we have chosen $v_0< v_1 < \infty$. So for $v>v_1$ and $u<u_0$, with $u_0<<-1$ : 
\begin{align*}
    0 &\leq \dfrac{1}{v-u}\int_{v_0}^{v_1} \dfrac{F}{\psi}\deriv v \leq \dfrac{1}{v-u}\int_{v_0}^{v_1} \deriv v\\
    0 & \leq \dfrac{1}{v-u}\int_{v_0}^{v_1} \dfrac{F}{\psi}\deriv v \leq \varepsilon
\end{align*}
For $v>v_1$ and $u<u_0$, with $v_1$ sufficiently large and $u_0<< -1$ we have, using \ref{prop_first_estimation}  $\vert F/\psi -1\vert \leq \varepsilon$ so the last term in \eqref{rtildeR} becomes:
\begin{align*}
    \left \vert \dfrac{1}{v-u}\int_{v_1}^v\left(\dfrac{F}{\psi}-1\right) \deriv v \right\vert & \leq   \dfrac{1}{v-u}\int_{v_1}^v\varepsilon \deriv v  \\
    &\leq \varepsilon \dfrac{v-v_1}{v-u} \leq \varepsilon\, .
\end{align*}
This concludes the proof. 

Let us now turn to $R \vert u \vert$. It is clearly non negative and in $\Omega_{u_0}$, we have $u = - s \tilde{r}$ with $s \in [0,1]$. Whence
\[ 0 \leq R \vert u \vert = s R \tilde{r} \leq R \tilde{r} < 1+ \varepsilon \, .\]
Finally, since $R$ is small in the neighbourhood of $i^0$ and $m$ is positive and bounded, we have trivially that $1-\varepsilon < 1-2m(u) R \leq 1$ in $\Omega_{u_0}$ provided $u <<-1$ is large enough in absolute value. \qed

\subsection{Proof of proposition \ref{prop_equivalence}} \label{ProofProp32}

We use \ref{prop_first_estimation}. Recall that

\begin{align*}
     \mathcal{E}_{\mathcal{H}_s}(\phi) =& \int_{\mathcal{H}_s} \left\lbrace u^2 \phi_u^2 + u^2 R^2 F(u,R) \phi_u \phi_R + \vert \nabla_{S^2}\phi\vert^2\left[u^2\dfrac{FR^2}{4s \psi}\left(2 + s(\psi - 1)\right) + (1+uR) \right] \right.\\
    & \left. + R^2 F  \left[\dfrac{2 + s(\psi - 1)}{4s \psi}\left( (2+uR)^2 - 2 m(u)u^2R^3\right) - (1+uR)\right]\phi_R^2 \right\rbrace \deriv u \wedge \deriv \omega \vert_{\mathcal{H}_s}
\end{align*}
\begin{enumerate}
     \item We begin with the term in front of $\vert \nabla_{S^2} \phi \vert^2$. We use the fact that $s = -u /\tilde{r}$
        \begin{equation*}
            \dfrac{u^2 R^2 F }{2s\psi} \left(2 + s(\psi - 1)\right) + (1 + uR) =  1 + \vert u \vert R \left[-1 + \dfrac{\tilde{r}RF}{2\psi} + \dfrac{s(\psi-1)\tilde{r}RF}{4\psi} \right]
        \end{equation*}
        \begin{equation*}
           -\frac{1}{2} - \varepsilon \leq \left[-1 + \dfrac{\tilde{r}RF}{2\psi} + \dfrac{s(\psi-1)\tilde{r}RF}{4\psi} \right] \leq -\frac{1}{2} + \varepsilon 
        \end{equation*}

        \begin{equation}
        \frac{1}{2} - \varepsilon \leq 1 + \vert u \vert R \left[-1 + \dfrac{\tilde{r}RF}{2\psi} + \dfrac{s(\psi-1)\tilde{r}RF}{4\psi} \right] \leq 1 \label{terme_angulaire}
        \end{equation}

    \item Now we study the term in front of $\phi_R^2$ that we denote by $f_{RR}$. 
    \begin{align*}
    f_{RR} = & R^2 F \left[\dfrac{2 + s(\psi -1)}{4 s \psi}\left((2+uR)^2 - 2m(u)u^2 R^3\right) - (1+uR)\right]\\
    =&\dfrac{R}{\vert u \vert} \left[ \left(\dfrac{\tilde{r}RF}{2\psi} + \dfrac{R^2F(\psi - 1)}{4 \psi}\right) \left((2+uR)^2 - 2m(u) u^2 R^3\right) - \vert u \vert R F(1+uR)\right] \\
    \simeq & \dfrac{R}{2\vert u \vert} \left( 3(\vert u \vert R)^2 - 6\vert u \vert R + 4\right)
    \end{align*} 
Let $x = \vert u \vert R$ and $f(x) = 3x^2 - 6x + 4$. If $x\in [0,1]$n then $1 \leq f(x) \leq 4$, hence, 
\begin{equation}
   \left( \dfrac{1}{2} - \varepsilon \right) \dfrac{R}{\vert u \vert} \leq f_{RR} \leq (2+\varepsilon ) \dfrac{R}{\vert u \vert} \label{terme_RR}
\end{equation}

    \item Finally we have to estimate the coefficient in front of $\phi_u\phi_R$. This term is exactly the same as for Schwarzschild and so we use similar arguments to \cite{mason_nicolas_2009}
    
    \begin{align}
    \left\vert R^2 u^2 F(u,R) \phi_u\phi_R \right\vert \leq& \left(R\vert u \vert\right)^{3/2} \vert u\phi_u \vert \left\vert \sqrt{\dfrac{R}{\vert u \vert}} \phi_R\right\vert \nonumber\\
    &\leq \dfrac{1}{2}\left(\lambda^2 u^2 \phi_u^2 + \dfrac{1}{\lambda^2}\dfrac{R}{\vert u \vert}\phi_R^2 \right), \; \lambda \in \mathbb{R}^{*} \label{EstimCrossTerm}
    \end{align}
And for an appropriate choice of $\lambda \in \mathbb{R}$ ($\lambda = 3/2$ for instance):
\begin{equation} \label{Constraintslambda}
    \dfrac{\lambda^2}{2} < 1,\;\; \dfrac{1}{2\lambda^2}< \dfrac{1}{2} -\varepsilon  .
\end{equation}
%
%
 \end{enumerate}
Using \eqref{terme_angulaire}, \eqref{terme_RR}, \eqref{EstimCrossTerm} and \eqref{Constraintslambda}, we obtain :
\begin{equation*}
    \mathcal{E}_{\mathcal{H}_s}(\phi) \simeq \int_{\mathcal{H}_s} \left[u^2\vert \partial_u\phi\vert^2  + \dfrac{R}{\vert u \vert}\vert\partial_R\phi\vert^2 + \vert \nabla_{S^2} \phi\vert^2 \right]\deriv u \wedge \deriv \omega \vert_{\mathcal{H}_s} \qed
\end{equation*}

\subsection{Proof of proposition \ref{prop_error_term}} \label{ProofProp33}

Let us recall the expression of the error term \eqref{ErrTermPhiFinalForm}.

\begin{equation}
    \Err(\phi)= \left[\left( -u^2 R^3 m'(u) + 2mR^2 (3+uR) \right) \phi_R^2 -2mR\phi \left( u^2 \phi_u - 2(1+uR)\phi_R\right)\right] \dfrac{(\tilde{r} R)^2}{\varphi \vert u \vert} \, .
\end{equation}

Using \ref{prop_first_estimation},
\begin{align*}
    \vert \Err(\phi) \vert
    \leq & \left[ \vert m'(u) \vert u^2 R^3 \vert \phi_R \vert ^2 + 2mR^2 \vert 3 + uR \vert \phi_R \vert^2 + 2mu^2 R \vert \phi \vert  \vert \phi_u \vert +4mR\vert 1 + uR \vert \vert \phi \vert \vert \phi_R \vert \right] \dfrac{\vert \tilde{r}R \vert^2}{\varphi \vert u \vert}\\
    \lesssim & \left[ \dfrac{R}{\vert u \vert} \vert \phi_R \vert^2 + \dfrac{R^2}{\vert u \vert} \vert \phi_R \vert^2 + \vert \phi \vert \vert \phi_u \vert + \dfrac{R}{\vert u \vert} \vert \phi \vert \vert \phi_R \vert \right]\\
    \lesssim & \dfrac{R}{\vert u \vert} \vert \phi_R \vert^2 + u^2 \vert \phi_u \vert^2 + \phi^2
\end{align*}
where we obtain the last line using the facts that $R <1$ and $1< \vert u \vert$ in $\Omega_{u_0}$. \qed

\subsection{Proof of Theorem \ref{thm_peeling_ho}} \label{AppendixProofThmPeeling}

First observe that the estimates on $\nabla_{S^2}^k \phi$ in the first part of the theorem are direct consequences of the fundamental estimates and the spherical symmetry of the spacetime. To prove the other estimates, we need to understand what new terms appear each time we apply $\partial_u$ and $\partial_R$ to the wave equation and how we can control them. We have :

\begin{equation} \label{ET}
\nabla^{(a}\left(T^{b)} T_{ab}(\Psi)\right)= \left( -u^2 R^3 m'(u) + 2mR^2 (3+uR) \right) \Psi_R^2 +\square_{\hat{g}}\Psi \left( u^2 \Psi_u - 2(1+uR)\Psi_R\right)
\end{equation}

where $\Psi$ is any smooth function on $\Omega_{u_0}$. In the following estimates, we shall take $\Psi = \partial_u^l \partial_R^k \phi$. When we integrate $\eqref{ET}$ on $\Omega_{u_0}$, we shall split the $4$-volume measure as $\deriv s \wedge (\nu \lrcorner \mathrm{dVol}^4 )$. Recall from the proof of the fundamental estimates that :
\begin{equation*}
    \nu \lrcorner \dvol \vert_{\mathrm{H}_s}\simeq \dfrac{1}{\vert u \vert} \deriv u \wedge \deriv \omega
\end{equation*}
So we shall need to control on each $\mathcal{H}_s$ the error term
\[ \Err (\Psi ) := \nabla^{(a}\left(T^{b)} T_{ab}(\Psi)\right) \frac{1}{\vert u \vert} \, .\]
A first immediate estimate is:
\[ 
    \vert \Err(\Psi) \vert \lesssim \dfrac{R}{\vert u \vert} \vert\Psi_R\vert^2  + \vert \square_{\hat{g}}\Psi \vert \left( \vert u \vert \vert \Psi_u \vert +\dfrac{1}{\vert u \vert}\vert \Psi_R \vert \right) \, . \]
The first term is controlled by the energy density on $\mathcal{H}_s$ as well as the square of $\vert u \vert \vert \Psi_u \vert$. But this is not the case for the square of $\dfrac{1}{\vert u \vert}\vert \Psi_R \vert$. Indeed, we do not have in $\Omega_{u_0}$ that $1/\vert u \vert ^2\leq R/\vert u \vert$ since, $R$ is allowed to tend to zero independently of the behaviour of $u$. However, using the fact that $s = \vert u \vert/\tilde{r}$, we have :
 \begin{align*}
     \dfrac{1}{\vert u \vert} \vert \Psi_R \vert = &\dfrac{1}{\sqrt{s}}\sqrt{\dfrac{R}{R\tilde{r} \vert u \vert}} \vert \Psi_R \vert \\
     \simeq & \dfrac{1}{\sqrt{s}}\sqrt{\dfrac{R}{\vert u \vert}} \vert \Psi_R \vert \, .
 \end{align*}
 So we obtain :
 \begin{align*}
    \vert \Err(\Psi) \vert  \lesssim \dfrac{1}{\sqrt{s}} \left( \left\vert \square_{\hat{g}}\Psi\right\vert^2 + \vert u \vert^2 \vert \Psi_u \vert^2 + \dfrac{R}{\vert u \vert}\vert \Psi_R\vert^2 \right) 
\end{align*}

Now, with the expression \eqref{energy_H_s} of the energy on $\mathcal{H}_s$ and using the fact that $s^{-1/2}$ is an integrable function near the origin, we can use Grönwall's lemma to prove the theorem, provided we know how to deal with the term $\vert \square_{\hat{g}}\Psi\vert^2$.

\subsubsection{Control of the wave operator}
 
 Now let us consider $\Psi = \partial_R^k \partial_u^l \phi$ where $\phi$ is a smooth solution of \eqref{equation_ondes} with compactly supported data on $\mathcal{H}_1$. 

The control of $\square_{\hat{g}} \Psi$ in this case is the main difference between Vaidya and Schwharzschild spacetimes. Indeed for Schwarzschild, we have a commutation relation between $\partial_u$ and the d'Alembertian operator, due to the fact that the metric was static :
 
 \begin{equation*}
     \left[ \partial_u, \square_{\hat{g}}\right]_{\mathrm{Sch}} = 0
 \end{equation*}
 
Now we have a d'Alembertien that depends both on $u$ and $R$ (see \eqref{d_alembertien_wave-operator}). And so :

 \begin{eqnarray}
    \left[ \partial_R , \square_{\hat{g}} \right] &= & -2R(1-3mR) \partial_R^2 - 2(1-6mR)\partial_R \label{derivee_square_R}\\
    \left[ \partial_u , \square_{\hat{g}} \right] &= & 2m'(u) R^3 \partial_R^2 + 6m'(u) R^2 \partial_R \label{derivee_square_u}
\end{eqnarray}    

The method to control $\square_{\hat{g}} \partial_R^k \partial_u^l \phi$ is the following :
\begin{enumerate}
    \item We begin by computing $\square_{\hat{g}} \phi_u$ and $\square_{\hat{g}} \phi_R$ : 
        \begin{align}
            \square_{\hat{g}} \phi_u = & -2m'(u) R^3 \partial_R \phi_R - 6m'(u)R^2 \phi_R - 2m'(u) R\phi - 2m(u)R\phi_u \label{dalembertien_phi_u}\\
            \square_{\hat{g}}\phi_R = & 2(1-7mR)\phi_R +2R(1-3mR)\partial_R \phi_R - 2m\phi  \label{dalembertien_phi_R}
        \end{align}
    \item For the sake of clarity and simplicity, we decompose our study into 3 parts : the case where $\Psi = \partial_R^k \phi$, the case where $\Psi = \partial_u^l \phi$ and the mixed case where $\Psi = \partial_u^l \partial_R^k \phi$. In each case, $\square_{\hat{g}} \Psi$ will involve some derivatives of the form $\partial_R^m \partial_u^n \phi$ with $m \leq k$ and $n \leq l$ multiplied by bounded smooth coefficients. These terms when squared can be estimated by the $L^2$ norm of $\partial_R^m \partial_u^n \phi$ which is in turn controlled by the energy of $\partial_R^m \partial_u^n \phi$. So all we need to consider are the terms involving derivatives of $\phi$ of order $k+l$ or above. In this manner, we shall prove that it is possible to control $\vert \square \partial_u^l \partial_r^k \phi\vert^2$ as we have done for the fundamental estimates. 
\end{enumerate}

\paragraph{Derivatives with respect to R}

We start with $\Psi = \partial_R^2 \phi$. We obtain $\square \Psi$ by applying $\partial_R$ to \eqref{dalembertien_phi_R}. The only terms of degree higher than $2$ are twice the same term
\[ 2R (1-3mR) \partial_R^3 \phi \]
coming once from the right-hand side of \eqref{dalembertien_phi_R} and once from the commutation of $\partial_R$ with $\square_{\hat{g}}$ (see \eqref{derivee_square_R}) in the left-hand side.
By induction, for $\Psi = \partial_R^k \phi$, the only terms of order larger than $k$ will be
\[ 2kR (1-3mR) \partial_R^{k+1} \phi \, . \]
We can estimate the square of this term as follows :

\[ 
        \left\vert 2kR(1-3mR) \partial_R^{k+1} \phi\right\vert^2 \lesssim  \; R^2 \left\vert \partial_R^{k+1}\phi\right\vert^2 
        \lesssim \; \dfrac{R^2 \vert u \vert}{\vert u\vert }\left\vert \partial_R^{k+1}\phi\right\vert^2 \simeq \dfrac{R}{\vert u \vert} \left\vert \partial_R^{k+1}\phi\right\vert^2 \, .
\]

We can state as an intermediary result that for the derivatives with respect to $R$ we have, as in the Schwarzschild case :

 \begin{equation*}
     \int_{\mathcal{H}_s} \left\vert \Err(\partial_R^k \phi)\right\vert \deriv u \wedge \deriv^2 \omega \lesssim \sum_{p=0}^k  \mathcal{E}_{\mathcal{H}_s}\left(\partial_R^p \phi\right)
 \end{equation*}
 
 \paragraph{Derivatives with respect to u}
 
 Now we set $\Psi = \partial_u^l \phi$. To obtain the wave operator of $\Psi$ we apply $\partial_u^{l-1}$ to \eqref{dalembertien_phi_u}. Similarly to what happened for the $\partial_R$ case, the only term involving derivatives of degree higher than $l$ is
 \[ -2lm'(u) R^3 \partial_R^2 \partial_u^{l-1} \phi \, .\]
 We estimate it as follows :
 
 \[
     \left\vert 2lm'(u) R^3 \partial_R^2 \partial_u^{l-1} \phi\right\vert^2 \lesssim R^6 \left\vert \partial_u^{l-1} \partial_R^2 \phi\right\vert^2
     \lesssim  \dfrac{R}{\vert u \vert} \left\vert \partial_u^{l-1} \partial_R^2 \phi\right\vert^2
 \]
 and this is controlled by the energy density on $\mathcal{H}_s$ of $\partial_u^{l-1} \partial_R^2 \phi$.
 
 So we can control this term with the energy of a derivative of order $l$ of $\phi$, however we notice that we need to introduce a control on derivatives in $R$ in order to control derivatives in $u$.
 
 \paragraph{Mixed case}
 
 Now we set $\Psi = \partial_u^l  \partial_R^k \phi$. We apply $\partial_R^{k}\partial_u^l$ to \eqref{equation_ondes} and the only terms involving derivatives of order higher than $k+l$ are similar to the terms we found in the previous two cases, namely
 \[ 2kR (1-3mR) \partial_u^l \partial_R^{k+1} \phi -2lm'(u) R^3 \partial_u^{l-1}\partial_R^{k+2} \phi \, .\]
 They are estimated just as before : 
 
 \begin{align*}
     \left\vert 2kR(1-3mR)\partial_u^l \partial_R^{k+1}\phi - 2lm'(u)R^3 \partial_u^{l-1}\partial_R^{k+2} \phi\right\vert^2 \lesssim\; & R^2 \left\vert \partial_u^l \partial_R^{k+1} \phi \right\vert^2 + R^6 \left\vert \partial_u^{l-1}\partial_R^{k+2}\phi \right\vert^2 \\
     \lesssim\; & \dfrac{R}{\vert u \vert} \left( \left\vert \partial_u^l \partial_R^{k+1} \phi \right\vert^2 + \left\vert \partial_u^{l-1}\partial_R^{k+2}\phi \right\vert^2 \right)
 \end{align*}
 which is in turn controlled by the sum of the energy densities on $\mathcal{H}_s$ of $\partial_u^l \partial_R^k \phi$ and $\partial_u^{l-1}\partial_R^{k+1}\phi$.
Hence we have the following estimate :
 
 \begin{equation*}
     \int_{\mathcal{H}_s}  \Err(\partial_u^l \partial_R^k \phi) \deriv u \deriv \omega \vert_{\mathcal{H}_s} \lesssim \frac{1}{\sqrt{s}} \sum_{p+q \leq k+l} \mathcal{E}_{\mathcal{H}_s}(\partial_R^p \partial_u^q \phi) \, .
 \end{equation*}
 
The theorem then follows from a Grönwall estimate. $\square$

\end{appendix}
\newpage
\nocite{*}
\printbibliography
\end{document}